\documentclass[%
prd%
,twocolumn%
 ,secnumarabic%
,amssymb, amsmath,nobibnotes, showpacs,showkeys]{revtex4}
\usepackage{bm}%
\usepackage{graphicx}
\expandafter\ifx\csname package@font\endcsname\relax\else
 \expandafter\expandafter
 \expandafter\usepackage
 \expandafter\expandafter
 \expandafter{\csname package@font\endcsname}%
\fi

\begin{document}

\title{The Immirzi parameter from an external scalar field}%

\author{Francesco Cianfrani$^{1}$, Giovanni Montani$^{123}$}%
\email{francesco.cianfrani@icra.it, montani@icra.it}
\affiliation{$^{1}$ICRA-International Center for Relativistic Astrophysics, Dipartimento di Fisica (G9),\\
Universit\`a  di Roma, ``Sapienza'', Piazzale Aldo Moro 5, 00185 Roma, Italy.\\ 
$^{2}$ENEA C.R. Frascati (Dipartimento F.P.N.), Via Enrico Fermi 45, 00044 Frascati, Roma, Italy.\\
$^{3}$ICRANet C. C. Pescara, Piazzale della Repubblica, 10, 65100 Pescara, Italy.}
\date{July 2009}%


\begin{abstract}

We promote the Immirzi parameter to be a minimally coupled scalar field and we analyzed the Hamiltonian constraints in the framework of Loop Quantum Gravity without the time gauge. Proper SU(2) connections can be defined and a term containing derivatives of the field $\beta$ enters into their definition. Furthermore, boost degrees of freedom are non-dynamical, while the super-momentum constraints coincide with the scalar field case. Hence, the kinematical Hilbert space can be defined as for gravity in presence of a minimally coupled scalar field. Then, we analyzed the dynamical implications of this scenario and we outline how a dynamical relaxation to a non-vanishing vacuum expectation value is predicted, so recovering the standard Loop Quantum Gravity formulation. 

\end{abstract}

\maketitle



\section{Introduction} 
The quantization of the gravitational field via the Loop Quantum Gravity (LQG) approach \cite{revloop} allows to define a proper kinematical Hilbert space, where geometrical operators have discrete spectra \cite{discr}. Hence, even though there are still problems with the definition of the physical Hilbert space and with the description of semi-classical states, nevertheless LQG can be considered the most successful non-perturbative Quantum Gravity model.  

Indeed, already at the kinematical level a fundamental ambiguity arises, the Immirzi parameter $\gamma$ \cite{Imm}. Such a quantity enters into the definition of the SU(2) connections (Ashtekar-Barbero-Immirzi connections \cite{ABI}), which are the basic variables within this scheme. At the Lagrangian level, $\gamma$ is just a factor in front of the so-called Holst modification \cite{Ho96} of the Einstein-Hilbert action. This modification does not affect the classical dynamics in the vacuum and, in fact, the value of $\gamma$ can be changed by a canonical transformation. However, once the theory is quantized, $\gamma$ labels inequivalent quantum sectors \cite{RT98} and it enters into the spectra of geometric operators, so fixing the scale of the space discreteness.    

Although there have been efforts to remove this sort of ambiguity (for instance in Covariant Loop Quantum Gravity \cite{Al}), nevertheless the prevailing idea among the LQG community is to regard $\gamma$ as a new fundamental parameter, which for instance can be fixed from the issue of reproducing the black hole entropy \cite{ABCK98} given by the Bekenstein formula \cite{Be73}. In particular, a topological interpretation was proposed for $\gamma$ \cite{GOP}, in analogy with the $\theta$ sector of QCD \cite{theta}. 

Recently, it was shown \cite{merprl} that once such a parameter is promoted to be a field, the divergent contribution to the chiral anomaly can be reabsorbed by renormalization. This point is taken as an indication that the Immirzi parameter is a fundamental field. In this respect, the possible cosmological implications of an Immirzi scalar and pseudo-scalar field were addressed in \cite{TY} and \cite{mc}, respectively. In this scheme the kinetic term for the Immirzi field is inferred by solving the second Cartan structure equation. Nevertheless, it is not all clear whether this procedure can be implemented on a quantum level. This point was not addressed in \cite{merprl}, since the analysis was performed on a Lagrangian level, only. However, the main problem consists in the fact that proper SU(2) Gauss constraint has not been obtained in this framework, thus the whole quantization in terms of SU(2) holonomies and fluxes is very questionable.

The relevant achievement of this work consists in inferring proper SU(2) Gauss constraints when dealing with an Immirzi scalar field, whose kinetic and potential terms do not depend on the geometry. We apply the formalism of LQG without the time gauge developed in \cite{prl,tgnnm}. In particular, we performed the Hamiltonian analysis, which outlines that the Gauss constraints of the Lorentz group are modified by the presence of the Immirzi field. Then, we solve second-class constraints arising in the Hamiltonian formulation and we promote boost degrees of freedom $\chi_a$ to be dynamical variables. Hence, we demonstrate that proper SU(2) Gauss constraints arise, while conjugate momenta to $\chi_a$ vanish. Furthermore from the analysis of the super-momentum constraints we conclude that the kinematical sector coincides with that one of gravity in presence of a scalar field in the time gauge. Therefore, the geometry and
the Immirzi field can be quantized simultaneously. This result confirms the ability of the adopted procedure to work out a proper set of constraints in view of the quantization. 

On a dynamical level, we emphasize the main differences with respect to the scalar field case. The most important achievement is the possibility to explain the relaxation of the Immirzi field to a fixed value, once spatial gradients are neglected. Therefore, standard Loop Quantum Gravity with the Immirzi parameter is inferred
as a low energy description.


\section{Hamiltonian structure}\label{1}

In LQG the starting point of the Lagrangian formulation for gravity is the Holst action \cite{Ho96}, which reads as follows (in units $c=8\pi G=1$) 
\begin{eqnarray}  
S_{Holst}=\int \sqrt{-g}e^\mu_A e^\nu_B R_{\mu\nu}^{CD}(\omega^{FG}_\mu){}^\beta\!p^{AB}_{\phantom1\phantom2CD}d^4x,
\label{act}\end{eqnarray}

$g_{\mu\nu}$ being the metric tensor, while $e^A_\mu$ and $\omega^{AB}_\mu$ denote 4-bein vectors and spinor connections, respectively.  The expression for $R^{AB}_{\mu\nu}$ is given by
\begin{equation}
R^{AB}_{\mu\nu}=2\partial_{[\mu}\omega^{AB}_{\nu]}-2\omega^A_{\phantom1C[\mu}\omega^{CB}_{\nu]}.
\end{equation}

The Immirzi parameter $\gamma=1/\beta$ is contained into $
{}^\beta\!p^{AB}_{\phantom1\phantom2CD}=\delta^{AB}_{\phantom1\phantom2CD}-\frac{\beta}{2}\epsilon^{AB}_{\phantom1\phantom2CD}$. 

In view of removing this ambiguity, we promote $\beta$ to a scalar field minimally coupled to gravity. We add a canonical kinetic term and a potential $V(\beta)$, such that the full action reads
\begin{eqnarray}  
S=\int \sqrt{-g}\bigg[e^\mu_A e^\nu_B R_{\mu\nu}^{CD}(\omega^{FG}_\mu){}^\beta\!p^{AB}_{\phantom1\phantom2CD}+\nonumber\\+
\frac{1}{2}g^{\mu\nu}\partial_\mu\beta\partial_\nu\beta-V(\beta)\bigg]d^4x.\label{actnm}
\end{eqnarray}

We denote by ${}^\beta\!\pi_{AB}^i={}^\beta\!p^{CD}_{\phantom1\phantom2AB}\pi_{CD}^i,$ and ${}^\beta\!\pi$ conjugate momenta to $\omega^{AB}_i$ and $\beta$, respectively, and we take $\{\omega^{AB}_i,\pi_{CD}^j,\beta,{}^\beta\!\pi\}$ as coordinates in the phase space.

By a Legendre transformation, the Hamiltonian density is obtained and it turns out to be a linear combination of the following constraints: 

\begin{eqnarray}   
H=\pi^i_{CF}\pi^{jF}_{\phantom1D}{}^\beta\!p^{CD}_{\phantom1\phantom2AB}R^{AB}_{ij}+\frac{1}{2}{}^\beta\!\pi^2+\nonumber\\+
\frac{1}{4}\pi^i_{AB}\pi^{jAB}\partial_i\beta\partial_j\beta+hV(\beta)=0,\label{sh}\\
H_i={}^\beta\!p_{AB}^{\phantom1\phantom2CD}\pi^j_{CD}R^{AB}_{ij}+{}^\beta\!\pi\partial_i\beta=0,\label{sm}\\
G_{AB}=\partial_i\pi^i_{AB}-2\omega_{[A\phantom2i}^{\phantom1C}\pi^i_{|C|B]}+\nonumber\\+
\frac{\beta\partial_i\beta}{(\beta^2+1)}\left(\pi^i_{AB}-\frac{1}{2\beta}\epsilon_{AB}^{\phantom{12}CD}\pi^i_{CD}\right)=0;\label{LG}\\
C^{ij}=\epsilon^{ABCD}\pi_{AB}^{(i}\pi_{CD}^{j)}=0,\label{C} \\ D^{ij}=\epsilon^{ABCD}\pi^k_{AF}\pi^{(iF}_{\phantom1\phantom2B}D_k\pi^{j)}_{CD}=0,
\label{D}
\end{eqnarray}

Here, $H$ and $H_i$ stand for the super-Hamiltonian and the super-momentum, respectively, while $G_{AB}=0$ are the constraints associated with the local Lorentz invariance. 

Other constraints $C^{ij}=0$ and $D^{kl}=0$ makes the whole system second-class and they are not associated with any gauge symmetry. It is worth noting that even though $\beta$ is promoted to a field, no modification occurs into the expression of $C^{ij}$ and $D^{ij}$ with respect to the vacuum case. However, $G_{AB}$ change and their expressions in terms of $\pi^i_{AB}$ do not reproduce anymore the Gauss constraints of the Lorentz group. This point affects significantly the analysis of constraints with respect to the cases discussed in \cite{prl,tgnnm} .

\section{Reduction to a first-class set of constraints}
Let us now denote $\pi^i_{0a}$ by $\pi^i_a$ and introduce their inverses $\pi^b_j$. It is convenient to express the 3-metric as $h_{ij}=-\frac{1}{\pi}T^{-1}_{ab}\pi^a_i\pi^b_j$, $\pi$ being the determinant of $\pi^a_i$, while $T^{-1}_{ab}$ contains some arbitrary space-time functions $\chi_a$ as follows, $T^{-1}_{ab}=\eta_{ab}+\chi_a\chi_b$. We point out that, as in the case of a non-minimally coupled scalar field \cite{tgnnm},  ${}^\beta\!h_{ij}$ does not coincide with the spatial metric $h_{ij}$, since one has ${}^\beta\!h_{ij}=(1+\xi\beta^2)h_{ij}$.

The constraints $C^{ij}=0$ and $D^{ij}=0$ are solved by restricting to the following hypersurfaces
\begin{eqnarray}
\pi^i_{ab}=2\chi_{[a}\pi^i_{b]},\label{scon1}\\ \omega^{\phantom1b}_{a\phantom1i}={}^\pi\!\omega^{\phantom1c}_{a\phantom1i}T^{-1b}_c+\chi_a\omega^{0b}_{\phantom{12}i}+\chi^b
(\omega_{a\phantom1i}^{\phantom10}-\partial_i\chi_a)+{}^1\!\omega^{\phantom1b}_{a\phantom1i},\label{scon2}
\end{eqnarray}

where ${}^\pi\!\omega^{\phantom1b}_{a\phantom1i}=\frac{1}{\pi^{1/2}}\pi^b_l{}^3\!\nabla_i(\pi^{1/2}\pi^l_a)$. The functions $\chi_a$ label different solutions and it can be demonstrated that the geometrical scenario underlying such a picture is characterized by the following set of 4-bein vectors 
\begin{eqnarray}
e^0=Ndt-\chi_a E^a_idx^i,\qquad e^a=E^a_iN^idt+E^a_idx^i.
\end{eqnarray} 

Hence, $\chi_a$ give the velocity components of the local Lorentz frame with respect to spatial hypersurfaces.

As for ${}^1\!\omega^{\phantom1b}_{a\phantom1i}$ the following relations hold

\begin{equation}
{}^1\!\omega^{\phantom1b}_{a\phantom1i}\pi^i_cT_b^{\phantom1c}=\frac{\beta\partial_i\beta}{\beta^2+1}\pi^i_a.
\end{equation}

In particular the explicit expression of ${}^1\!\omega^{ab}_{\phantom1i}=\eta^{ac}{}^1\!\omega^{\phantom1b}_{c\phantom1i}$ reads
\begin{eqnarray}
{}^1\!\omega^{ab}_{\phantom1i}=T^{[a}_c\bigg(-\frac{2(1+\chi^2)^2}{\chi^4+2\chi^2+2}\eta^{b]d}+\nonumber\\+
\frac{2+\chi^2}{\chi^4+2\chi^2+2}\chi^{b]}\chi^d\bigg)\pi^c_i\pi_d^j\frac{\beta\partial_j\beta}{\beta^2+1}.\label{mods}
\end{eqnarray}

This additional term is the modification with respect to case in which a scalar field is present \cite{tgnnm}. It does not contribute to $D^{ij}$, while as soon as the relations (\ref{scon1}), (\ref{scon2}) are inserted into $G_{AB}$ one gets 
\begin{equation}
G_{0a}=\chi^bG_{ab}. \label{bred}
\end{equation}

The relations above insure that the constraints associated to boosts become redundant. This feature will allow us to define a canonical transformation mapping the initial phase space variables to coordinates on the hypersurfaces  
defined by conditions (\ref{C}) and (\ref{D}). 

Furthermore, the presence of terms containing derivatives of the field $\beta$ into spinor connections $\omega_{\phantom1i}^{ab}$ is not surprising. In fact such terms are predicted also in a first-order formulation for the action (\ref{act}) when solving the II Cartan structure equation. 

\section{Kinematical sector}\label{sk}

The analysis of the relic set of constraints is complicated by the fact that a non-trivial symplectic form is induced by the conditions (\ref{scon1}) and (\ref{scon2}). We simplifies such an analysis by looking for a set of conjugate variables parameterizing the associated hypersurfaces into the phase space. Furthermore, the requirement of dealing with a canonical transformation mapping the initial set of variables to the final ones will replace the constraints $G_{0a}=0$.

Having in mind the standard LQG formulation, we define generalized Ashtekar-Barbero-Immirzi variables $\widetilde{A}^a_i$ such that their conjugate momenta are inverse densitized triads, which we denote by $\widetilde{\pi}^i_a$. The expression of $\widetilde{\pi}^i_a$ is the following one 
\begin{equation}
\widetilde{\pi}^i_a=S^b_a\pi^i_b,\qquad S^a_b=\sqrt{1+\chi^2}\delta^a_b+\frac{1-\sqrt{1+\chi^2}}{\chi^2}\chi_a\chi_b,
\end{equation}

while $\widetilde{A}^a_i$ is given by
\begin{eqnarray}
\widetilde{A}_i^a=S^{-1a}_b\bigg((1+\chi^2)T^{bc}\left(\omega_{0ci}+{}^\pi\!D_i\chi_c\right)-\frac{\beta}{2}\epsilon^b_{\phantom1cd}({}^\pi\!\omega^{cf}_{\phantom1\phantom2i}\nonumber\\
T^{-1d}_{\phantom1f}+{}^1\!\omega^{cd}_{\phantom{12}i})+\beta\frac{2+\chi^2-2\sqrt{1+\chi^2}}{2\chi^2}\epsilon^{abc}\partial_i\chi_b\chi_c\bigg),\quad\label{ABI}
\end{eqnarray}

where ${}^\pi\!D_i\chi_c=\partial_i\chi_c-{}^\pi\!\omega_{c\phantom1i}^{\phantom1b}\chi_b-\frac{1}{1+\chi^2}{}^1\!\omega_{c\phantom1i}^{\phantom1b}\chi_b$. 

Let us now write $G_{ab}=0$ in terms of a $\{\widetilde{A}_i^a,\widetilde{\pi}^i_a\}$, using the condition \begin{equation}
\partial_i\pi^i_a-{}^\pi\!\omega_{a\phantom1i}^{b}\pi^i_b-\frac{\partial_i\chi^2}{2(1+\chi^2)}\pi^i_a=0.
\end{equation}

We find that it is possible to infer the Gauss constraints of the SU(2) group, but the following re-scaling has to be performed
\begin{eqnarray}
{}^{(\beta)}\!\widetilde{\pi}^i_a=\beta\widetilde{\pi}^i_a\qquad {}^{(\beta)}\!\widetilde{A}_i^a=\frac{1}{\beta}\widetilde{A}_i^a.
\end{eqnarray}

Therefore, from $G_{ab}=0$ we get
\begin{equation}    
G_a=\partial_i{}^{(\beta)}\!\widetilde{\pi}^i_a+\epsilon_{ab}^{\phantom{12}c}{}^{(\beta)}\!\widetilde{A}^b_i{}^{(\beta)}\!\widetilde{\pi}_c^i=0.\label{gcs2}
\end{equation}

Hence, \emph{the structure of a SU(2) gauge theory can be found out also in presence of an Immirzi-like field}.
This is a decisive result in view of applying the LQG quantization procedure. 

As mentioned above, other constraints are inferred from the requirement that the transformation $\{\omega_i^{AB},{}^\beta\!\pi^j_{CD},\beta,{}^\beta\!\pi\}\rightarrow\{{}^{(\beta)}\!\widetilde{A}^a_i,\chi_b,{}^{(\beta)}\!\widetilde{\pi}_c^j,\pi^d,\beta,{}^{(\beta)}\!\widetilde{\pi}\}$ is canonical. It is rather impressive that the conditions associated to the request of dealing with a canonical transformation coincide simply with the vanishing behavior of the conjugate momenta to $\chi_a$, {\it i.e.}
\begin{equation}
\pi^a=0.\label{pc}
\end{equation}

Therefore, \emph{variables $\chi_a$ do not play any dynamical role} and henceforth they behave as Lagrangian multipliers.

In addition the conjugate momentum to $\beta$ does not coincide anymore with the original one ${}^\beta\!\pi$. This is due to the presence of $\beta$ into ${}^{(\beta)}\!\widetilde{A}_i^a$ and ${}^{(\beta)}\!\widetilde{\pi}^i_a$. Hence the new conjugate momentum is given by
\begin{equation}
{}^{(\beta)}\!\widetilde{\pi}={}^\beta\!\pi-\frac{1}{\beta}{}^{(\beta)}\!\widetilde\pi^i_a\left({}^{(\beta)}\!\widetilde{A}_i^a+\frac{1}{2}\epsilon^a_{\phantom1bc}{}^{(\beta)}\!\widetilde{\omega}^{bc}_i\right),\label{mtr}
\end{equation}

where 
the explicit expression of ${}^{(\beta)}\!\widetilde{\omega}^{bc}_i$ is 
\begin{equation}{}^{(\beta)}\!\widetilde{\omega}^{\phantom1b}_{a\phantom2i}={}^{(\beta)}\!\widetilde{\pi}^b_j{}^{(\beta)}\!\nabla_i{}^{(\beta)}\!\widetilde{\pi}^j_a,\end{equation}

${}^{(\beta)}\!\nabla_i$ being the covariant derivative with respect to ${}^{(\beta)}\!h_{ij}$.

In terms of these variables the super-momentum constraints coincide with those ones of a scalar field, {\it i.e.}
\begin{equation}
H_i={}^{(\beta)}\!\widetilde{\pi}_a^j{}^\beta\!\widetilde{F}^a_{ij}+{}^{(\beta)}\!\widetilde{\pi}\partial_i\beta,\label{sms}
\end{equation}

${}^{(\beta)}\!\widetilde{F}^a_{ij}$ being $2\partial_{[i}{}^{(\beta)}\!\widetilde{A}^a_{j]}+\epsilon^a_{\phantom1bc}{}^{(\beta)}\!\widetilde{A}^b_{i}{}^{(\beta)}\!\widetilde{A}^c_{j}$. 

Therefore, from constraints (\ref{gcs2}), (\ref{pc}) and (\ref{sms}) one infers that \emph{the kinematical sector of LQG without the time gauge in presence of the Immirzi field coincides with that one of LQG in presence of a scalar field}.

\section{Dynamical content}

The analysis of the super-Hamiltonian constraint emphasizes that the dynamics of the Immirzi field does not coincide with the analogous one for a minimally coupled scalar field. 

In particular, the expression of $H$ in the adopted set of variables is the following one
\begin{eqnarray}
H=\frac{{}^{(\beta)}\!\widetilde{\pi}^i_a{}^{(\beta)}\!\widetilde{\pi}^j_b}{2}\epsilon^{\phantom{12}c}_{ab}{}^{(\beta)}\!\widetilde{F}^c_{ij}-\frac{(\beta^2+1)}{\beta^2}{}^{(\beta)}\!\widetilde{\pi}^i_a{}^{(\beta)}\!\widetilde{\pi}^j_b\nonumber\\
\big(\partial_{[i}{}^{(\beta)}\!\omega^{ab}_{\phantom{12}j]}-{}^{(\beta)}\!\omega^{ac}_{\phantom{12}[i}{}^{(\beta)}\!\omega^{\phantom1b}_{c\phantom1j]}\big)+\frac{1}{2}{}^{\beta}\!\pi^2-\nonumber\\-
\beta{}^{\beta}\!h{}^{\beta}\!h^{ij}{}^{(\beta)}\!\widetilde{\nabla}_i\partial_j\left(\frac{1}{\beta}\right)+
\frac{{}^{\beta}\!h{}^{\beta}\!h^{ij}}{2}\partial_i\beta\partial_j\beta+{}^{\beta}\!h\frac{V}{\beta^3}.\label{shf}
\end{eqnarray} 
 
It is worth noting the presence of ${}^\beta\!\pi$, which is not the conjugate momenta to $\beta$, since it also contains terms involving ${}^{(\beta)}\!\widetilde{A}^a_{i}$ and ${}^{(\beta)}\!\widetilde\pi^i_a$ (\ref{mtr}). Therefore, within this scheme a non-trivial interaction is predicted between $\beta$ and geometric degrees of freedom, thus making the dynamics of this field a tantalizing subject of investigation in LQG.

A further modification is given by the presence of terms with second spatial spatial derivatives of the field $\beta$. 

On a kinematical level, geometrical variables ${}^{(\beta)}\widetilde{\pi}^i_a$ describe the fictitious re-scaled spatial metric ${}^{\beta}\!h_{ij}=\beta h_{ij}$ and not the true one $h_{ij}$. For this reason, ${}^{\beta}\!h_{ij}$ behaves as the basic geometric variable and this idea will be confirmed by the quantum analysis of the next section.

Hence, the effective potential for $\beta$ is given by $V_{eff}=V(\beta)/\beta^3$. If we assume $V(\beta)$ to be a quartic potential, {\it i.e.}  $V(\beta)=\mu^2\beta^2+\frac{1}{4}\lambda\beta^4$, we finally get the following expression
\begin{equation}
V_{eff}(\beta)=\frac{\mu^2}{\beta}+\frac{\lambda}{4}\beta.\label{pot}
\end{equation}

Such a potential exhibits a non vanishing minimum.

Hence neglecting spatial gradients and the interaction with the geometry, a dynamical relaxation to a non-vanishing expectation value is predicted for the Immirzi field. This relaxation is able to explain its parametric role in standard LQG. In fact, at energy scales much smaller that the minimum of $V_{eff}(\beta)$ the main contribution to the dynamics is given by the vacuum expectation value $\beta_{min}$, while the oscillations of $\beta$ around such a minimum behave as massive perturbations.   

In particular, the resulting Immirzi parameter reads as 
\begin{equation}
\beta^2_{min}=4\frac{\mu^2}{\lambda}.
\end{equation}

Hence, \emph{the introduction of the Immirzi field minimally coupled with the geometry allows to reproduce as a low energy effect LQG with the Immirzi parameter}.  

\section{Loop quantization of the model.}\label{4}

As outlined in section \ref{sk}, the kinematical constraints coincide with those ones of gravity in presence of a scalar field, a part from the conditions (\ref{pc}). Hence, as soon as any dependence from $\chi_a$ is avoided for quantum states, the kinematical Hilbert space $\textsc{H}_{Kin}$ can be defined as in standard LQG in presence of a scalar field $\beta$ \cite{Thmatt,KLB,Thold}.  

In particular, the field $\beta$ is quantized by virtue of point-like holonomies \cite{Thmatt,KLB}, {\it i.e.}
\begin{equation}
U_x(\beta)=e^{i\lambda(x)\beta(x)},
\end{equation}

and of volume-integrated momenta $\Pi(V)=\int_V {}^{(\beta)}\!\widetilde{\pi} d^3x$. This choice of variables allows to define a background-independent measure $d\mu_U$. 

As for gravitational degrees of freedom, they can be treated by the standard LQG procedure \cite{revloop}, based on the algebra generated by holonomies of the SU(2) connections ${}^{(\beta)}\!\widetilde{A}_i^a$ along edges, 
and fluxes of momenta across surfaces $S$.

The resulting Hilbert space is given by 
\begin{equation}
\textsc{H}_{Kin}=L_2(\bar{A},d\mu_0^{SU(2)})\otimes L_2(\bar{U},d\mu_U)
\end{equation}

where $\bar{A}$ and $\bar{U}$ denote the algebra of distributional connections and distributional scalar fields, respectively. The proper SU(2)-invariant sector can be find out for $L_2(\bar{A},d\mu_0^{SU(2)})$ and basis vectors are invariant spin-networks $T_\alpha$. Because $\beta$ does not carry SU(2) quantum numbers, no SU(2)-invariant space has to be select out. As a consequence, unlike the case in which a Higgs field is present, point-like holonomies are not restricted on spin-network vertexes.  

However, ${}^{(\beta)}\!\widetilde{\pi}_a^i$ are inverse densitized vectors of the re-scaled fictitious metric ${}^{\beta}\!h_{ij}$. Hence the area operator of a surface $S$ with respect to the fictitious geometry ${}^{\beta}\!h_{ij}$ acts on a spin network $T_\alpha$ along a graph $\alpha$ as follows
\begin{equation}
{}^{(\beta)}\!A(S)T_\alpha({}^{(\beta)}\!\widetilde{A})=\sum_e l_p^2\sqrt{j_e(j_e+1)} h_e({}^{(\beta)}\!\widetilde{A}),\label{area}
\end{equation}

$j_e$ being the quantum numbers of the $SU(2)$ representation associated with the edge $e$, while the sum runs on all the non-tangential edges of $\alpha$ starting from $S$. In the expression above we re-insert the constants $G$, $\hbar$ and $c$, which combine to form the Planck length $l_P$.

True geometrical operators turn out to depend on the field $\beta$. The representation of the associated operator on a quantum level is complicated, because only point like holonomies are well defined (see \cite{Thold} for a proper regularization in the case of a Higgs-like scalar field) and $\beta$ enters by a negative power. If we can define a state in which $1/\beta$ acquires the expectation value $<1/\beta>=1/\beta_0(x)$ in some region containing $S$, then the expectation value of the true area operator (associated with the metric $h_{ij}$) becomes 
\begin{equation}
<A(S)>_\alpha=\sum_e \frac{l_p^2}{\beta_0(p_e)}\sqrt{j_e(j_e+1)}, 
\end{equation}

$p_e$ being the point of the edge $e$ which belongs to $S$.

Therefore, in this case the contribution which the field $\beta$ provides to the area spectra is the expected factor due to the redefinition of ${}^{(\beta)}\!\widetilde{\pi}^i_a$. However, since now $\beta$ is a fundamental field and not a parameter, such a contribution qualifies ${}^{\beta}\!h_{ij}$ as the basic geometric structure.  

We point out that in the limit in which LQG is fully recovered, {\it i.e.} for a non-vanishing vacuum expectation value of the Immirzi field, $\beta$ is constant and our analysis of the area spectrum overlaps the case treated in \cite{AC} when the non-minimally coupled scalar field is constant on the horizon of a black hole.

\section{Conclusions}
We discussed the dynamical implications of promoting the Immirzi parameter to a scalar field with canonical kinetic and potential terms. We demonstrated that the analysis of constraints allows to define proper SU(2) connections, which are canonically conjugated to re-scaled densitized 3-bein vectors of the spatial metric. It was emphasized that the time-gauge condition is not at all necessary since, like in vacuum \cite{prl} and in presence of a non-minimally coupled scalar field \cite{tgnnm}, conjugate momenta to variables $\chi_a$ are constrained to vanish. Hence, the kinematical sector of such a model reproduced exactly that one of a scalar field minimally coupled to geometry, so the development of the kinematical Hilbert space can be carried on as in \cite{Thmatt}. 

Indeed, the dynamical behavior of $\beta$ was characterized by two main differences with respect to a minimally coupled scalar field. The first peculiarity consisted in the appearance of second spatial derivatives of the Immirzi field into the super-Hamiltonian constraint. The second difference is due to the the presence of $\beta$ into the definition of SU(2) connections. Such a feature implied that the momenta canonically conjugated to $\beta$ needed a redefinition and a non-trivial interaction with geometric variables was predicted. 

Therefore the dynamics of the Immirzi field deserves further investigations. In this respect the application of this scheme to Loop Quantum Cosmology is intriguing, since the Immirzi field would be a distinctive component of the cosmological bath or it would manifest a peculiar behavior as a clock-like matter field.

Then, we suggested to regard the re-scaled metric ${}^{\beta}\!h_{ij}$ as the basic geometric quantity. Such an interpretation came from the fact that ``electric field'' variables ${}^{(\beta)}\!\widetilde{\pi}^i_a$ are densitized inverse 3-bein of ${}^{\beta}\!h_{ij}$. Hence, an effective potential was predicted for $\beta$ which was able to explain the dynamical relaxation to a non-vanishing vacuum expectation value. This feature was required to find out standard LQG with the Immirzi parameter as a low energy effect.    

We did not discuss the quantization of the full Hamiltonian constraint (\ref{shf}). The treatment of inverse powers of the field $\beta$ is highly non-trivial and the definition of a proper super-Hamiltonian operator requires the extension of the techniques developed in \cite{Thold}.

\end{document}